\documentclass[twocolumn,secnumarabic,amssymb, nobibnotes, aps, pra]{revtex4-1}

\usepackage{framed}
\usepackage{amsfonts}
\usepackage{amsmath}
\usepackage{graphicx}
\usepackage{color}

\usepackage[hidelinks]{hyperref}
\usepackage{xcolor}
\hypersetup{
	colorlinks,
	linkcolor={blue!80},
	citecolor={blue},
	urlcolor={blue!80!black}
}

\begin{document}
\title{Parametric localized patterns and breathers in dispersive quadratic cavities}
\author{P. Parra-Rivas$^{1,2}$, C. Mas-Arab\'i$^1$, F. Leo$^1$}

\affiliation{ 
	$^1$Service OPERA-photonics, Université libre de Bruxelles, 50 Avenue F. D. Roosevelt, CP 194/5, B-1050 Bruxelles, Belgium\\	
	$^2$Laboratory of Dynamics in Biological Systems, KU Leuven Department of Cellular and Molecular Medicine, University of Leuven, B-3000 Leuven, Belgium\\}

\date{\today}


\begin{abstract}
	
	We  study the formation of localized patterns arising in doubly resonant dispersive optical parametric oscillators. They form through the locking of fronts connecting a continuous-wave and a Turing pattern state. This type of localized pattern can be seen as a slug of the pattern embedded in a homogeneous surrounding. They are organized in terms of a {\it homoclinic snaking} bifurcation structure, which is preserved under the modification of the control parameter of the system. We show that, in the presence of phase mismatch, localized patterns can undergo oscillatory instabilities which make them breathe in a complex manner.
\end{abstract}
\maketitle

\section{Introduction}


The formation of temporal dissipative structures in passive nonlinear resonators is currently attracting renewed attention.
	These robust nonlinear attractors correspond, in the spectral domain, to optical frequency combs, i.e. coherent light sources composed of a set of equidistant spectral lines that find applications in metrology and spectroscopy \cite{diddams_evolving_2010}.
	While mostly generated in Kerr cavities at first \cite{leo_temporal_2010-1,herr_temporal_2014-1}, their emergence in quadratically nonlinear cavities has been intensely studied in past few years because they offer advantages that may help overcome some of the limitations of Kerr combs \cite{ricciardi_optical_2020}.
	Indeed, quadratic combs allow to address spectral regions far from that of the continuous wave pump and may facilitate self-referencing.

Dissipative structures can be either localized or extended \cite{cross_pattern_1993}  and their dynamics and shape determine the breadth and coherence of the underlying comb \cite{parra-rivas_dynamics_2014-1}. Localized dissipative structures, hereafter LSs, are particularly attractive as they correspond to denser combs. 
	The LSs arising in Kerr cavities are very well known \cite{gomila_bifurcation_2007,parra-rivas_dynamics_2014,godey_stability_2014,parra-rivas_dark_2016,parra-rivas_bifurcation_2018}. Bright and dark temporal cavity solitons for example have been shown to correspond to ultra-coherent frequency combs \cite{herr_temporal_2014-1,xue_mode-locked_2015-1}.
	Conversely, the bifurcation structure of quadratic resonators are still poorly understood.
	
LSs have been first studied in quadratic diffractive cavities, where they form within the transverse plane to the propagation direction of light \cite{etrich_solitary_1997-1,oppo_formation_1994,trillo_stable_1997,staliunas_localized_1997,staliunas_spatial-localized_1998,oppo_domain_1999,oppo_characterization_2001,staliunas_transverse_2003,taranenko_patterns_2000}. In this context solitary waves appear in second harmonic enhanced quadratic cavities \cite{etrich_solitary_1997-1}, and domain walls (DWs) between two plane wave states exist and interact in optical parametric oscillators (OPOs), leading to the formation of more complex states and dynamics \cite{oppo_formation_1994,trillo_stable_1997,staliunas_localized_1997,longhi_localized_1997,staliunas_spatial-localized_1998,oppo_domain_1999,oppo_characterization_2001,staliunas_transverse_2003,taranenko_patterns_2000}.

LSs have been also recently investigated in dispersive quadratic cavities, where they arise along the propagation direction. In this situation such states have been studied in cavity-enhanced second-harmonic generation \cite{hansson_quadratic_2018,villois_soliton_2019}, in degenerate OPO (DOPO) with pure quadratic nonlinearity \cite{parra-rivas_frequency_2019,parra-rivas_localized_2019}, and in the context of competing nonlinearities \cite{villois_frequency_2019}.

In all these studies LSs either consist in a solitary wave, i.e. a cavity soliton \cite{villois_frequency_2019,villois_soliton_2019}, or they form due to the locking of DWs connecting two different but coexisting continuous wave (CW) states \cite{hansson_quadratic_2018,parra-rivas_frequency_2019,parra-rivas_localized_2019}.
However, there is yet another mechanism that can lead to the formation of LSs: the locking of fronts or DWs connecting a subcritical Turing pattern and a stable CW state. In this context a LS consist in a portion of a pattern embedded in the CW state. Turing patterns emerge from a 
modulational instability (MI), and although its dynamical implications has been studied by different authors in this context \cite{leo_walk-off-induced_2016,leo_frequency-comb_2016,mosca_modulation_2018}, its potential for the generation of LSs has not been yet analyzed.
In this work, we focus on the study of parametric LSs emerging in dispersive doubly resonant DOPOs in the framework of this last mechanism. Our investigation reveals the presence of multi-stability between LSs of different widths, result of a particular bifurcation structure known as {\it homoclinic snaking} \cite{woods_heteroclinic_1999}. These states and its bifurcation structure persist through the modification of different parameters of the system, and, moreover, they can undergo instabilities which make them oscillate in a complex fashion.

The article is organized as follows. In section~\ref{sec:1} we briefly introduce the model and the methods that we apply in our study. Section~\ref{sec:2} is devoted to the linear stability analysis of the CW state and the emergence of Turing patterns. Later, in Sec.~\ref{sec:3}, we investigate the origin and formation of the LSs, analyze their bifurcation structure and stability, and classify the different dynamical regimes in a phase diagram. In Sec.~\ref{sec:4} the oscillatory dynamics of the LSs is studied, and their dynamical origin determined. Finally, in Sec.~\ref{sec:5} we present a short discussion and the conclusions.

\begin{figure}[t]
	\centering
	\includegraphics[scale=0.75]{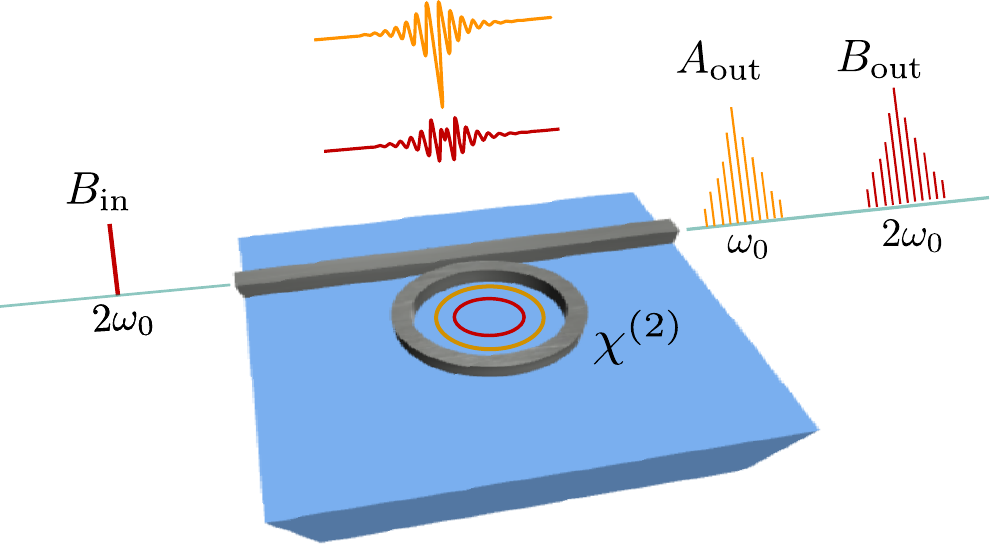}
	\caption{(Color online) Schematic example of a doubly resonant DOPO. A ring resonator with a $\chi^{(2)}$ nonlinearity is driven by a CW field $B_{in}$ at frequency $2\omega_0$. The quadratic interaction gives rise to a field $A$ with carrier frequency $\omega_0$ that resonates together with $B$, and therefore to two frequency spectra around $2\omega_0$ and $\omega_0$, respectively.}
	\label{sketch_DOPO}
\end{figure}

\section{The Model and Methods}\label{sec:1}
Here we consider a doubly resonant DOPO cavity as the one shown in Fig.~\ref{sketch_DOPO}. High finesse cavities of this type can be described by the mean-field model: 
\begin{subequations}\label{MF}
	\begin{equation}\label{MF1}
	\partial_t A=-(1+i\Delta_1)A-i\eta_1\partial_{x}^2A+i B\bar{A}
	\end{equation}
	\begin{equation}\label{MF2}
	\partial_t B=-(\alpha+i\Delta_2)B-\left(d\partial_{x}+i\eta_2\partial_{x}^2\right)B+i A^2 +S,
	\end{equation}
\end{subequations}
where $A$ and $B$ are the slowly varying envelopes of the normalized signal electric field centered at frequency $\omega_0$, and pump field centered at the frequency $2\omega_0$ \cite{parra-rivas_frequency_2019}.
In this context, $t$ corresponds to the normalized slow time describing the evolution of fields after every round-trip, and $x$ is the normalized fast time. The parameters $\Delta_{1,2}$ are the normalized cavity phase detunings, with $\Delta_2=2\Delta_1+\varrho$ and $\varrho$ the normalized wave-vector mismatch between the fields $A$ and $B$ over one roundtrip; $\alpha$ is the ratio of the round-trip losses $\alpha_{1,2}$ associated with the propagation of the signal and pump fields; $\eta_{1,2}$ are the group velocity dispersion (GVD) parameters of $A$ and $B$; $d$ is the normalized rate of temporal walk-off between both fields; and $S$ is the driven field amplitude or pump at frequency $2\omega_0$. Here $\eta_1=+1$($-1$) denotes normal (anomalous) GVD, and $\eta_2$ can take any positive or negative value. 

The model~(\ref{MF}) is formally equivalent to those describing diffractive spatial cavities \cite{oppo_formation_1994,staliunas_transverse_2003,zambrini_convection-induced_2005}, with $\eta_1\approx 2\eta_2$ where $\eta_j>0$ are the diffraction parameters and $x$ represents a transverse spatial dimension.

Stationary states of this system satisfy the set of ordinary differential equations:
\begin{subequations}\label{staMF}
	\begin{equation}\label{staMF1}
	-i\eta_1\partial^2_xA-(1+i\Delta_1)A+i B\bar{A}=0,
	\end{equation}
	\begin{equation}\label{staMF2}
	-\left(d\partial_{x}+i\eta_2\partial_{x}^2\right)B-(\alpha+i\Delta_2)B+i A^2 +S=0.
	\end{equation}
\end{subequations}
Notice that the steady states can also be studied in terms of a single integro-differential equation for $A$, as done in \cite{parra-rivas_frequency_2019,parra-rivas_localized_2019}.

To reveal the dynamics and the bifurcation structure of the different DSs circulating in this cavity we combine direct numerical simulations through a pseudo-spectral split-step scheme \cite{montagne_wound-up_1997} for the integration of Eqs.~(\ref{MF}), and numerical parameter continuation algorithms to track the steady states solutions of Eqs.~(\ref{staMF}) \cite{doedel_numerical_1991-1,doedel_numerical_1991}.


To calculate the stability of the different DSs we solve numerically the eigenvalue problem:
\begin{equation}\label{eigen}	
\mathcal{L}\psi=\sigma\psi,
\end{equation}
where $\sigma$ is the eigenvalue associated with the eigenmode $\psi$, and $\mathcal{L}$ represents the linear operator obtained from the linearization of Eq.~(\ref{MF}) around a given stationary state. A stationary DS is stable whenever Re$[\sigma]<0$ and unstable otherwise. 
Due to the periodic nature of the resonator we consider periodic boundary conditions in $A$ and $B$, and a domain size $L=70$.

In the following, we fix $\alpha=1$, $\eta_1=1$, $\eta_2=-0.25$, and $d=0$. The effect that the walk-off may have on the results presented here is beyond the scope of this work and will be presented elsewhere. 

Initially, in Secs.~\ref{sec:2} and \ref{sec:3} we consider perfect phase matching $(\varrho=0)$, and the implications of phase mismatch $(\varrho\neq0)$ on the LSs is presented in Sec.~\ref{sec:4}.

\section{Stability of the continuous wave states and periodic patterns} \label{sec:2}
The CW state solutions of the system are obtained by setting the derivatives in Eqs.~(\ref{staMF}) to zero, i.e. $(\partial_x^2A,\partial_xB,\partial_x^2B)=(0,0,0)$. Then, the CW states can be completely determined through the signal field $A$, and correspond to the trivial state $A_0=0$ and the non-trivial one $A=|A^\pm|e^{i\phi^\pm}$, with intensity  
\begin{equation}\label{hom}
|A^{\pm}|^2=\alpha(\Delta_1\Delta_2-1)\pm\sqrt{S^2-\alpha^2(\Delta_2+\Delta_1)^2},
\end{equation}
and phase 
\begin{equation}
\phi^\pm=\frac{1}{2}{\rm cos}\left[\frac{|A|^2+\alpha(1+\Delta_2^2)}{S\sqrt{1+\Delta_2^2}}\right].
\end{equation}
Then the $B$ component of the CW can be obtained through the equation 
\begin{equation}
B=\frac{iA^2+S}{\alpha+i\Delta_2}.
\end{equation}
The nontrivial state emerges from a pitchfork bifurcation in $A_0$ occurring at $S_p=\alpha\sqrt{(1+\Delta_1^2)(1+\Delta_2^2)}$. When $\Delta_2\Delta_1-1<0$, $A^+$ arises supercritically. In contrast, if $\Delta_2\Delta_1-1>0$, $A^-$ emerges subcritically, and eventually undergoes a fold at $S_t=\alpha(\Delta_1+\Delta_2)$, where it becomes $A^+$.

\begin{figure}[t]
	\centering
	\includegraphics[scale=0.95]{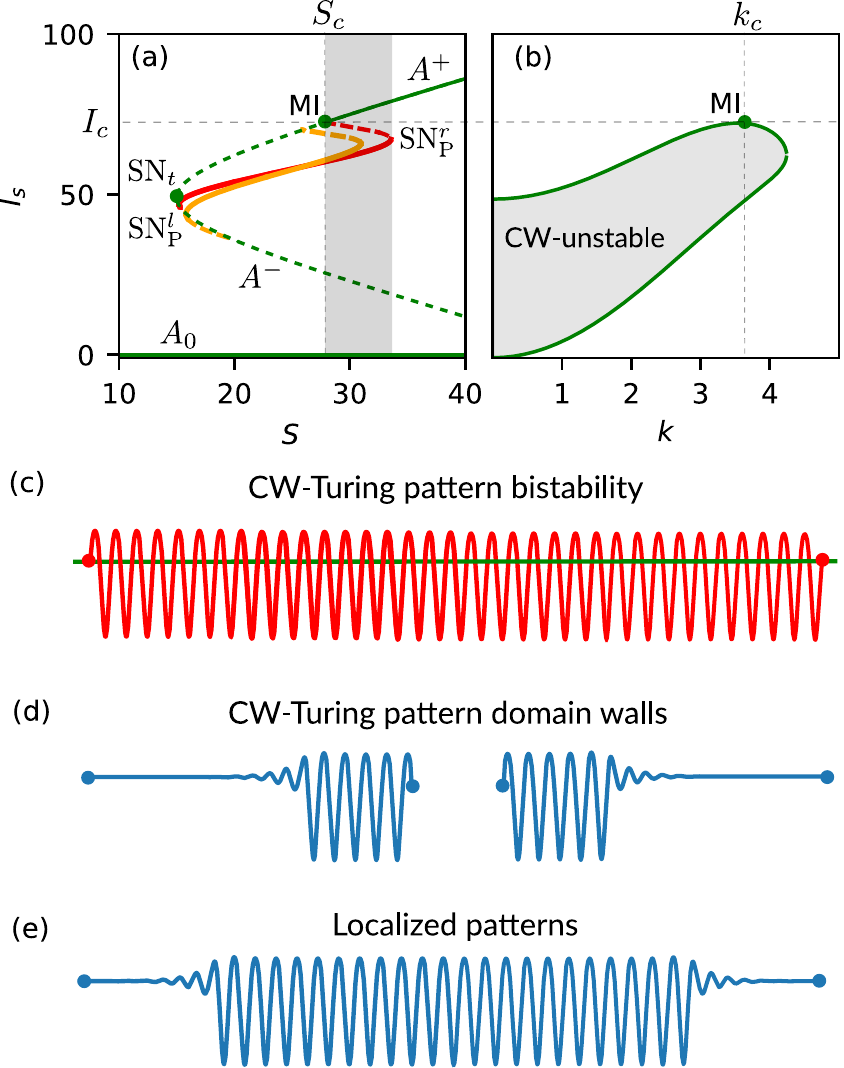}
	\caption{(Color online) Panel (a) shows the CW state for $\Delta_1=-5$ and (b) its marginal instability curve. Solid (dashed) lines represents stable (unstable) states. The gray area in panel (a) shows the interval in $S$ where the pattern $P$ and $A^+$ coexist and are stable (i.e. the bistability region). This coexistence is shown schematically in panel (c). In (d) two fronts connecting $A^+$ and the pattern are shown. Through the locking of these fronts localized patterns like the one shown in (e) arise.}
	\label{marginal}
\end{figure}

Figure~\ref{marginal}(a) shows a CW subcritical configuration for $\Delta_1=-5$ and $\varrho=0$. This bifurcation diagram shows the energy of $A$:
$$||A||^2=\frac{1}{L}\int_{-L/2}^{L/2}|A(x)|^2dx,$$
as a function of the pump intensity $S$. Note that for the CW states  $||A||^2=|A|^2\equiv I_s$.

At this stage we can calculate the linear stability of the CW against plane wave perturbations of the form $e^{\sigma t}\psi_k+c.c.$, where $\sigma$ is the growth rate of the perturbation and $\psi_k$ is the mode associated with the wavenumber $k$. This analysis amounts to solving the eigenvalue problem (\ref{eigen}), with $\mathcal{L}$ being evaluated at the CW state.

For the set of parameters considered in this work $A_0$, is stable for $S<S_p$ and unstable otherwise.

The stability of $A^{\pm}$ is characterized by the {\it marginal instability curve} (MIC) obtained from the condition Re$[\sigma]=0$ \cite{parra-rivas_localized_2019}. The MIC defines the band of unstable modes $\psi_k$ and is plotted in Fig.~\ref{marginal}(b) for the same parameter values than Fig.~\ref{marginal}(a). Thus, $A^{\pm}$ is unstable against a given mode $\psi_{k}$ if the intensity $I_s$ of CW lays inside the MIC [see gray region in Fig.~\ref{marginal}(b)], and stable otherwise. In correspondence, Fig.~\ref{marginal}(a) shows stable (unstable) CW states using solid (dashed) lines. While $A^-$ is always unstable, $A^+$ is only unstable between the fold SN$_t$ and the MI occurring at $S_c$. At this point the perturbation of the CW state slowly evolves to a periodic Turing pattern characterized by a critical wavenumber $k_c$. We refer to this pattern as the primary pattern $P$. Note that this instability corresponds to the maximum of the MIC, occurring at $(k,I_s)=(k_c,I_c)$.  

The primary pattern can be then tracked in the parameter $S$ applying numerical continuation algorithms, and as result one obtains the solution branches plotted in red in Fig.~\ref{marginal}(a). For our particular choice of parameters, the pattern arises subcritically from the MI at $S_c$ (i.e. unstably) and stabilizes in the saddle-node SN${_P}^r$. After that the periodic state remains stable until SN${_P}^l$, where it changes stability once more and eventually connects back to $A^{-}$ in a spatial resonance \cite{parra-rivas_bifurcation_2018-1}. Within the range of parameters studied in this work the primary pattern always arises subcritically.

The subcriticality of the pattern defines a hysteresis loop with the CW $A^+$, resulting in the coexistence of the pattern and the CW in a given interval of the pump intensity $S$. This bistability region is marked with a
gray box in Fig.~\ref{marginal}(a). Fig.~\ref{marginal}(c) shows schematically such coexistence, where the green line represent the CW $A^+$, and the subcritical pattern $P$ is plotted in red.

We need to point out that together with the primary pattern arising from the MI, there are many others that emerge all along $A^+$ and that connect back to $A^-$. A similar scheme has been described in detail in the context of Kerr cavities \cite{parra-rivas_bifurcation_2018-1}. The orange line Fig.~\ref{marginal}(a) shows one of such type of secondary patterns.

\section{Localized patterns: bifurcation structure and stability}\label{sec:3}

\begin{figure*}[t]
	\centering
	\includegraphics[scale=0.95]{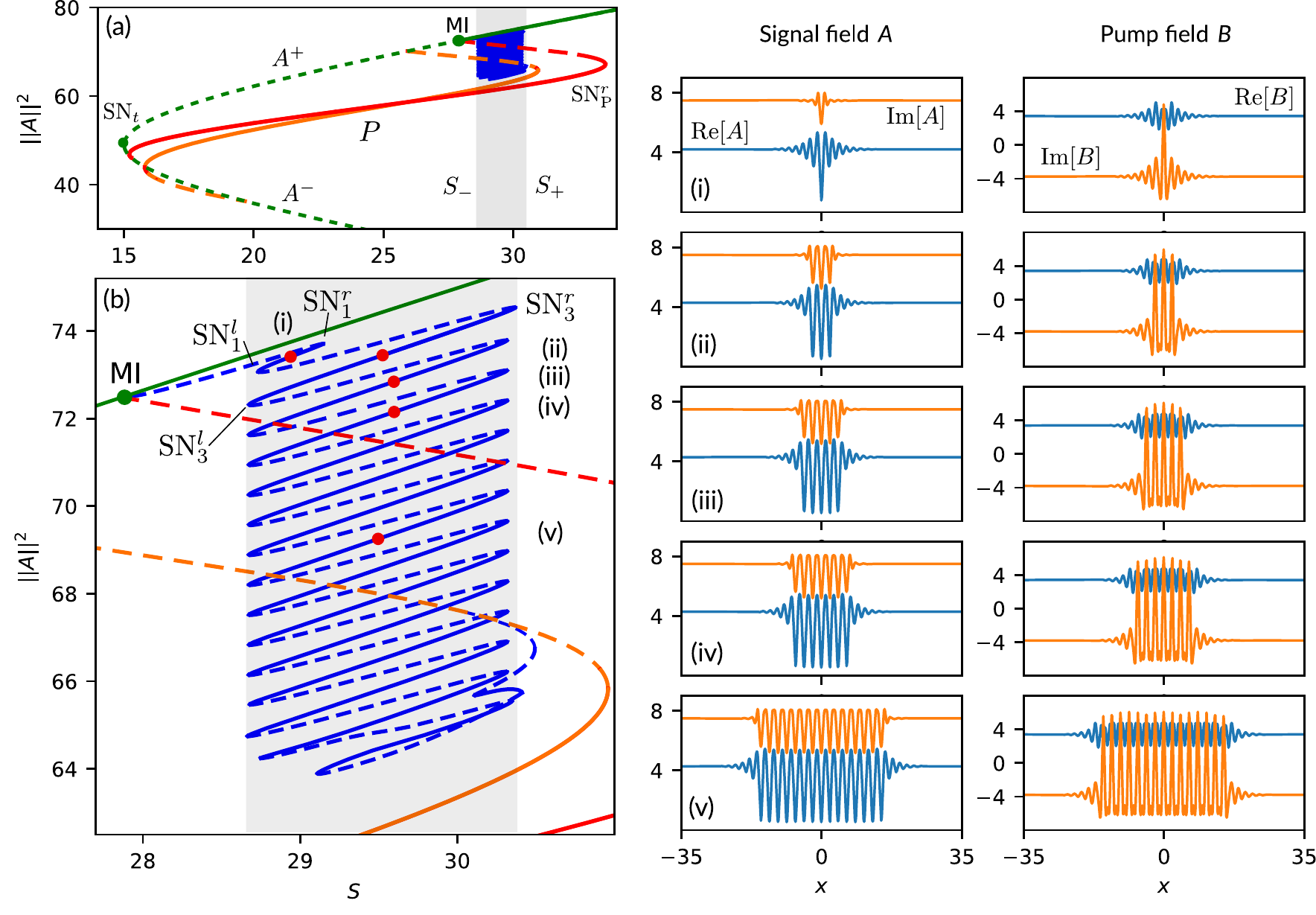}
		\includegraphics[scale=0.95]{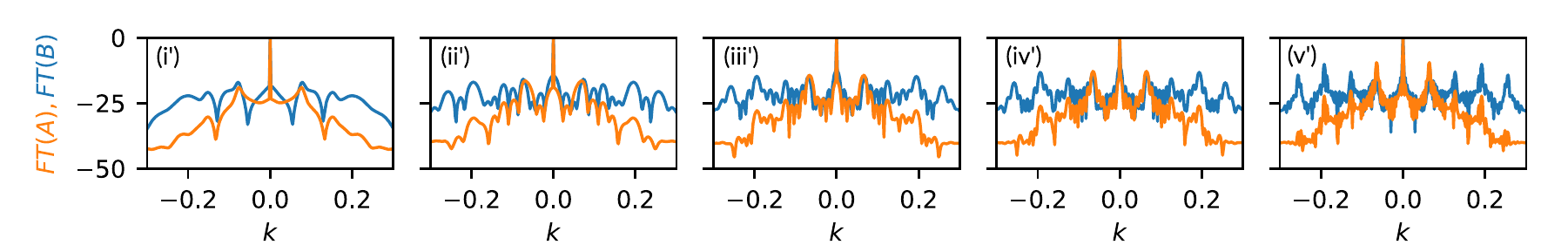}
	\caption{(Color online) Homoclinic snaking and LPs for $\Delta_1=-5$. Panel (a) shows the CW state, two pattern branches arising from it (red and orange lines), and the solution branches corresponding to the LPs (in blue). Solid (dashed) lines represent stable (unstable) states. The gray area within the interval $[S_{-},S_{+}]$ shows the pinning region. Panel (b) is a close-up view of (a) around the homoclinc snaking structure (blue set of branches). The red pattern branch $P$ represents the primary pattern emerging from the MI located at $(S_c,I_c)$; the orange line shows another periodic pattern arising from the CW from a point below $I_c$. The labels (i)-(v) stand for the LPs shown on the right. The first column shows the real and imaginary part of the signal field $A$, while the second column the pump field $B$. Panels (i')-(v') correspond to their frequency spectrum $FT(\cdot)=10{\rm log}_{10}(\mathcal{F}[|\cdot|^2])$ in orange and blue respectively. Here we consider $L=70$ and phase matching ($\varrho=0$).}
	\label{Dia_2}
\end{figure*}

Within the bistability region depicted in Fig.~\ref{marginal}(a), fronts or DWs like the ones shown in Fig.~\ref{marginal}(d) may arise, interact and eventually lock forming LSs of different widths, as the one plotted in Fig.~\ref{marginal}(e). This type of LS consists of a slug of the pattern state embedded in the CW $A^+$, and this is why they are normally called {\it localized patterns} (LPs). To properly understand the formation of these LPs, one has to approach the problem from a geometrical perspective as discussed in detail in \cite{woods_heteroclinic_1999,gomila_bifurcation_2007,makrides_existence_2019}.

The bifurcation structure associated with these type of states is shown in Fig.~\ref{Dia_2}(a)-(b) where we plot the energy $||A||^2$ as a function of $S$ for $\Delta_1=-5$. Panel (b) shows a close-up view of panel (a) that highlights the different branches of the structure. Representative LPs profiles along these branches are plotted in panels (i)-(v), where the first (second) column shows the real and imaginary part of $A$ ($B$), and their Fourier transform $\mathcal{F}$ [i.e. $FT(\cdot)=10{\rm log}_{10}(\mathcal{F}[|\cdot|^2])$] are plotted in panels (i')-(v'). To calculate these LPs and track them as a function of $S$ we have applied numerical continuation algorithms starting from a suitable initial guess. 

This bifurcation diagram, known as {\it homoclinic snaking}, consists in a sequence of LPs solution branches (see lines in blue) which oscillate back and forth within the {\it snaking or pinning region} defined by the parameter interval $S_-<S<S_+$ [see gray region in Fig.~\ref{Dia_2}]. This oscillation reflects the successive addition of a pair of pattern rolls, one on each side of the state, as one follows the diagram downwards (i.e. decreasing energy). Furthermore, these branches undergo a sequence of saddle-node bifurcations where the LPs gain or lose stability. We have labeled these bifurcations SN$_i^{l,r}$ where $i$ indicates the number of peaks in the LP and $l$ ($r$) stands for the left (right) side of the bifurcation diagram. In what follows we refer to a state with $i-$peaks as LP$_{i}$. 
Fig.~\ref{Dia_2}(b) shows that the limits of the snaking region (i.e. $S_-$ and $S_+$) are well determined by the positions of any saddle-node SN$_i^{l,r}$ with $i>1$, such as SN$_3^{l,r}$. The stability of these LSs is indicated using solid (dashed) lines for the stable (unstable) states [see Fig.~\ref{Dia_2}(a)-(b)]. The homoclinic snaking structure reflects the presence of multi-stability between LPs of different widths within the pinning region $S\in[S_-,S_+]$.

The LPs shown here are composed by an odd number of pattern peaks, and they emerge subcritically from the MI together with the primary pattern [see Fig.~\ref{Dia_2}(b)]. Together with these families of solutions there is another set of LPs, characterized by an even number of pattern peaks, which also undergo homoclinic snaking \cite{woods_heteroclinic_1999,burke_snakes_2007,parra-rivas_bifurcation_2018}. To avoid further confusion we do not show these states. 

In finite domains, the roll (i.e. peak) adding process must terminate at some point. Furthermore, in periodic domains like ours the snaking branches terminate on one of the many branches of periodic patterns that are present [see orange branches in Fig.~\ref{Dia_2})(a)-(b)]. 

We have to point out that this scenario is not intrinsic to the field of  nonlinear optics, but generic, appearing in many different context from hydrodynamics and material sciences to plant ecology \cite{ woods_heteroclinic_1999,burke_snakes_2007,mercader_convectons_2011,lo_jacono_magnetohydrodynamic_2011,beaume_homoclinic_2011,brena-medina_subcritical_2014,kreilos_fully_2017,zelnik_desertification_2017,gandhi_spatially_2018,parra-rivas_bifurcation_2018}.

\begin{figure}[t]
	\centering
	\includegraphics[scale=1]{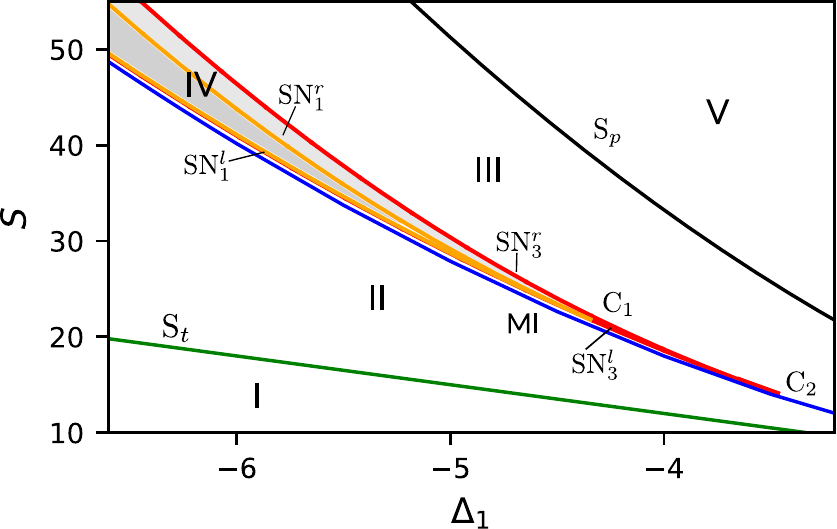}
	\caption{(Color online) Phase diagram in the $(\Delta_1,S)-$parameter space showing the main bifurcation lines of the system. The blue line corresponds to the MI, the black line $S_p$ represents the picthfork bifurcation of the CW states, and the green line $S_t$ corresponds to SN$_t$. The dynamical regions of the system are labeled as I-V and defined in the main text.}
	\label{Phase_1}
\end{figure}

At this point one may wonder how the snaking region, i.e. the region of existence of LPs, and shape of the states modify for different values of the cavity phase detuning $\Delta_1$.
To answer this question we perform a two-parameter continuation of SN$_1^{r,l}$ (which define the region of existence of LP$_{1}$) and SN$_3^{l,r}$ (which define the limit points $S_-$ and $S_+$) in the $(\Delta_1,S)-$parameter space. As a result, we obtain the phase diagram shown in Fig.~\ref{Phase_1} where SN$_3^{l,r}$ are plotted in red, and SN$_1^{l,r}$ in orange. The snaking region is shown in gray. 
For completeness we also plot the lines  $S_p$ and $S_t$ corresponding to the pitchfork and fold bifurcations of the CW state, and the MI (see blue line).

Decreasing $|\Delta_1|$ the snaking region shrinks: SN$_1^{l}$ and SN$_1^{r}$ approach each other and eventually collide in a cusp bifurcation C$_1$ occurring at $(S^{C_1},\Delta_1^{C_1})\approx(21.81,-4.3446)$, where LP$_1$ disappears. Decreasing $|\Delta_1|$ even further, SN$_3^r$ and SN$_3^l$ annihilate one another in a second cusp $C_2$ at $(S^{C_2},\Delta_1^{C_2})\approx(14.0762,-3.4696)$, where LP$_{3}$ disappears. In a similar way, a cascade of such collisions marks the successive destruction of the rest of LPs.  

At this stage of the study we can identify five different dynamical regimes [see Fig.~\ref{Phase_1}] that we describe as follows:
\begin{itemize}
	\item[I]: Only $A_0$ exist and is stable for pump intensities bellow $S_t$.
	\item[II]: The $A^+$ CW state is modulationally unstable within the MI at $S_c$ and $S_t$, and coexists with stable $A_0$. In this region Turing patterns exist.
	\item[III]: Between the MI and the pichfork $S_p$, $A^+$ is stable and coexists with the stable $A_0$. In this region one may expect the formation of LSs through the locking of DWs between $A_0$ and $A^+$, as described in \cite{parra-rivas_localized_2019}. 
	\item[IV]: This area corresponds to the snaking region (i.e. $S\in[S_-,S_+]$), where the system exhibit multi-stability of LPs. 
	
	\item[V]: In this region, $A^+$ and $A_0$ coexist although $A_0$ is unstable. 
\end{itemize}

Within the limits of region IV, LPs undergo standard homoclinic snaking. However, although LPs persist for larger values of $|\Delta_1|$, their bifurcation structure becomes much more complex, involving the formation of isolas, and the reconnection with solution branches of different unstable states.  
Furthermore, we have also verified that the LPs presented here and the LSs formed through the locking of domain walls studied in \cite{parra-rivas_localized_2019} can coexist and interact when modifying some of the control parameters. As a result one may expect the emergence of new types of LSs and bifurcation scenarios, which has been reported in other fields \cite{zelnik_implications_2018,ruiz-reynes_patterns_2020}. 
Nevertheless, the study of these complex scenarios is beyond the scope of the present paper, and will be presented in detail in another work.

\begin{figure}[t]
	\centering
	\includegraphics[scale=1.0]{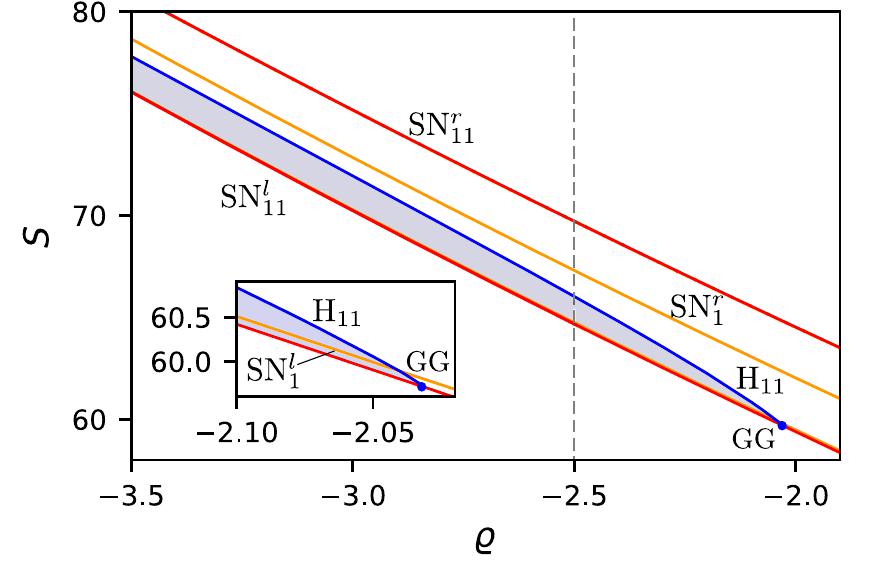}
	\caption{(Color online) Phase diagram in the $(\varrho,S)-$parameter space for $\Delta_1=-6$. The blue line corresponds to the Hopf bifurcation H$_{11}$, and the red and orange lines to SN$_{1}^{l,r}$ and SN$_{11}^{l,r}$ respectively. Within the gray region LP$_{11}$ undergoes a breathing behavior as depicted in Fig.~\ref{Breather}. The Hopf line arises from a codimension-two Gavrilov-Guckenhaimer (GG) bifurcation where the H$_{11}$ and SN$_{11}^{l}$ collide. The vertical dashed line corresponds to the diagram shown in Fig.~\ref{Dia_3}.}
	\label{Phase2}
\end{figure}

\section{Oscillatory dynamics in the presence of phase mismatch}\label{sec:4}
So far we have considered perfect phase matching between the signal and pump fields (i.e. $\varrho=0$). However, in practice, this condition is difficult to achieve, and non-negligible phase mismatch $\varrho$ between $A$ and $B$ may arise, which can influence the dynamics and stability of LSs \cite{villois_soliton_2019,erkintalo_dynamics_2019}. 

In this section we investigate how the LPs, their bifurcation structure, and stability are modified when varying $\varrho$. 
\begin{figure}[t]
	\centering
	\includegraphics[scale=0.9]{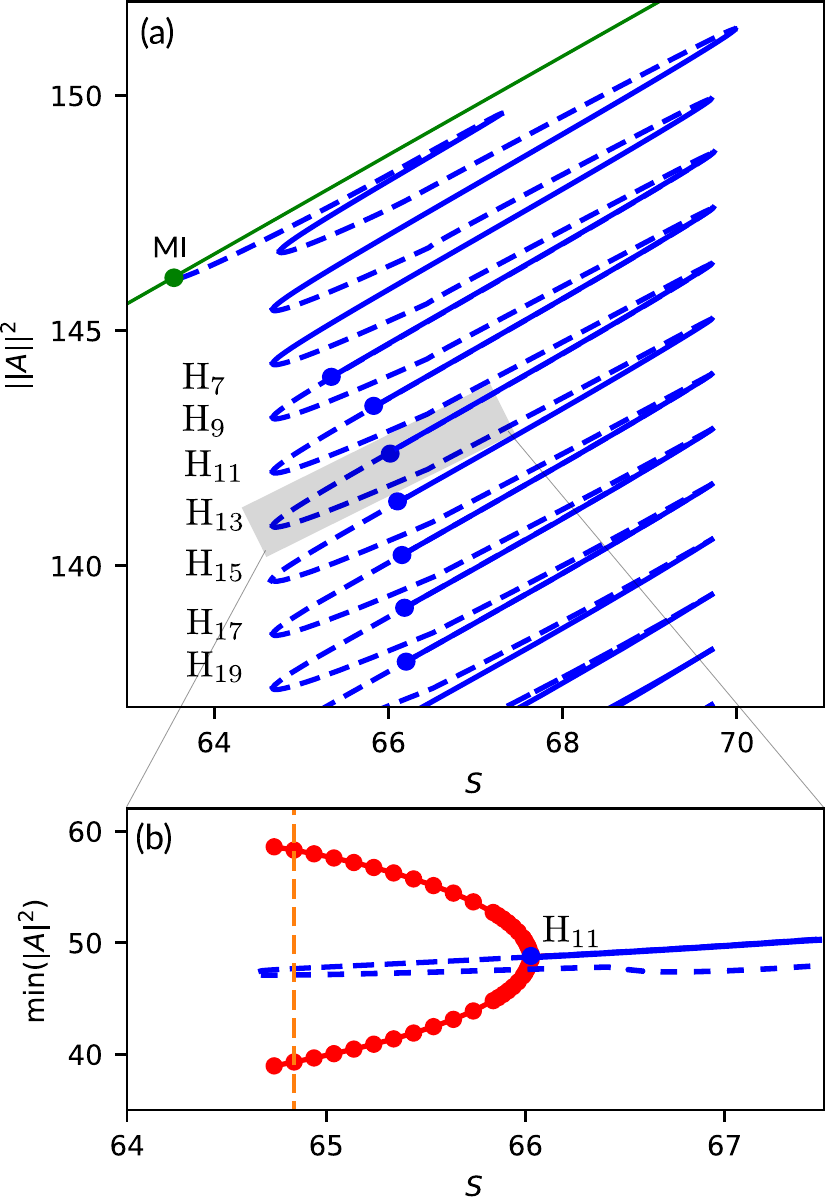}
	\caption{(Color online) (a) Homoclinic snaking diagram for $\varrho=-2.5$ and $\Delta_1=-6$. The Hopf bifurcations H$_i$ are marked using ${\color{blue}\bullet}$. Solid (dashed) lines correspond to stable (unstable) LPs. (b) shows the solution branches of a LP$_{11}$ using min$(|A|^2)$ [see gray box in (a)]. The extrema of the breather oscillations are plotted using ${\color{red}\bullet}$. The vertical dashed line corresponds to the LP$_{11}-$breather shown in Fig.~\ref{Breather}.}
	\label{Dia_3}
\end{figure}
The modification of the pinning region as a function of the phase mismatch is shown in the $(\varrho,S)$-phase diagram of Fig.~\ref{Phase2} for $\Delta_1=-6$, where SN$_1^{l,r}$ and $S_{\pm}$  are plotted using yellow and red solid lines (in particular here we identify $S_{\pm}$ with SN$_{11}^{l,r}$).  Increasing $|\varrho|$, the pinning region shifts to larger values of $S$, although it maintains its extension. Here we just show the range $[-3.5,-1.9]$. Note that these bifurcation lines continue all the way until $\varrho\geq0$.
\begin{figure*}[t]
	\centering
	\includegraphics[scale=1]{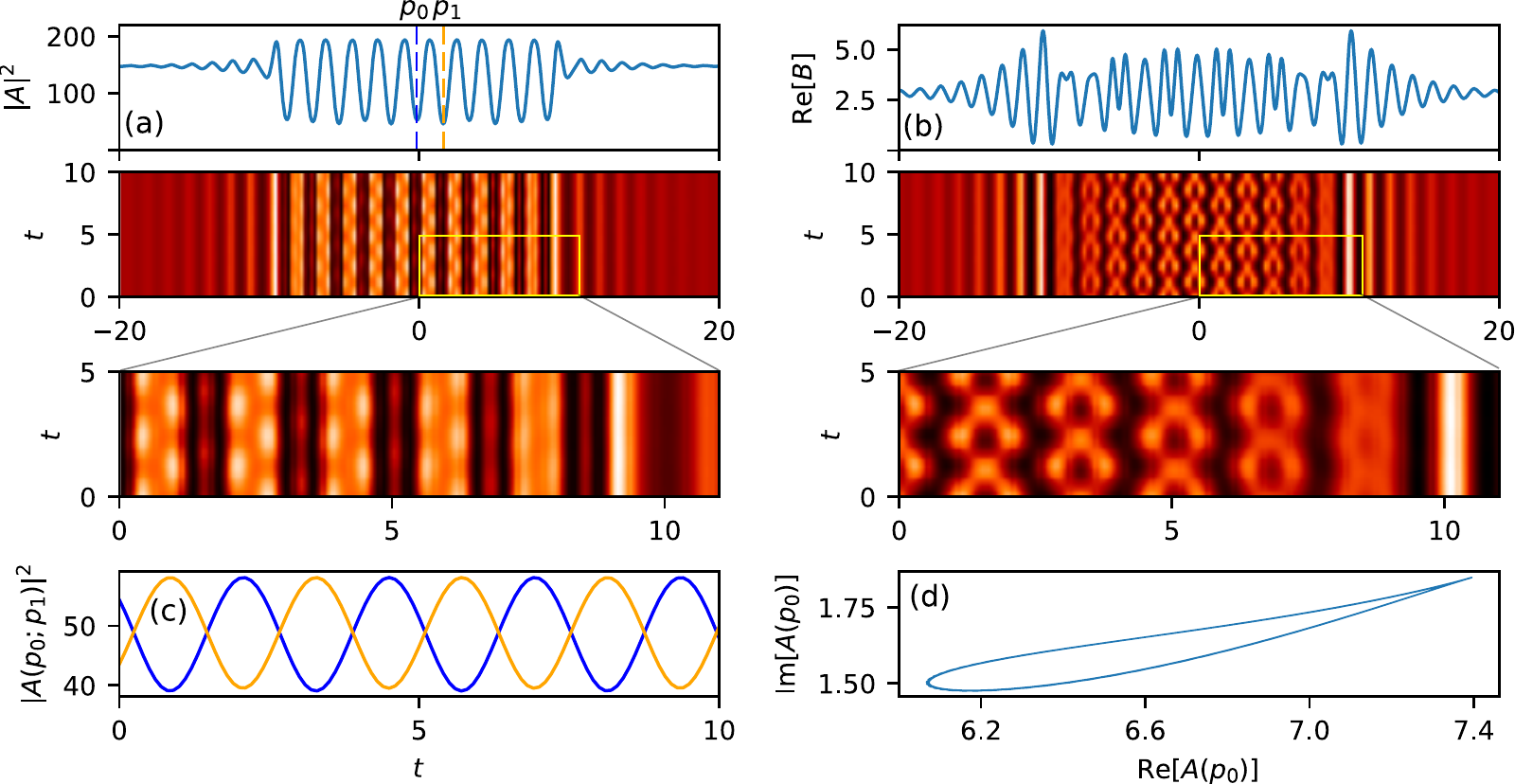}
	\caption{(Color online) LP$_{11}-$breather for $S=64.9$, $\Delta_1=-6$, and $\varrho=-2.5$. (a) shows the temporal evolution of the intensity of the signal field, i.e. $|A|^2$, for an interval of time $t=10$. The bottom panel shows an enlarged view of the oscillatory behavior, and the top panel shows the intensity profile of the LP at $t=10$. Panel (b) shows the temporal evolution of the real part of the pump field $B$ for the same interval of time. The top and bottom panels show the profile at time $t=10$, and the close-up view of the oscillatory behavior. Panel (c) shows the temporal evolution of the intensity of the peaks labeled $p_0$ and $p_1$ in (a). In panel (d) we show the limit cycle defined by the projection of the breather evolution on the variables Re$[A](p_0)$ and Im$[A](p_0)$.}
	\label{Breather}
\end{figure*}
Figure~\ref{Dia_3}(a) shows the homoclinic snaking corresponding to $\varrho=-2.5$ [see vertical dashed line in Fig.~\ref{Phase2}]. 
The stability analysis of these structures reveals that the previously stable LPs undergo a sequence of Hopf bifurcations where they become dynamically unstable to oscillatory LPs or LP-{\it breathers} like the one shown in Fig.~\ref{Breather}. These Hopf bifurcations are labeled with H$_i$, where $i$ represents the number of peaks of the LP. Proceeding downwards in energy, the first Hopf that appears is H$_7$ followed by H$_{9}$ to H$_{19}$. 
We can notice that, increasing the number of peaks, the H$_i$ bifurcations asymptotically reaches a single constant value of $S\approx66.2$. We suspect that this phenomenon may be related with a Hopf bifurcation undergone by the pattern, although the characterization of this accumulation needs further investigation.

In Fig.~\ref{Dia_3}(b) we show a detailed view of a portion of the bifurcation diagram of Fig.~\ref{Dia_3}(a) around H$_{11}$ [see gray box in (a)], where the minimum of $|A|^2$ is plotted against the pump intensity $S$. The red dots show the modification that the breather's central peak extrema undergo when decreasing the pump intensity $S$ from H$_{11}$.

An example of a LP$_{11}$-breather is shown in Fig.~\ref{Breather} for $S\approx64.9$ [see orange vertical dashed line in Fig.~\ref{Dia_3}(b)]. The colormaps in  panels (a)-(b) show the temporal evolution of the breather in $|A|^2$, and Re$[B]$ respectively, and on top of them we plot the profiles of the fields for $t=10$; the panels below (a) and (b) show a close-up view of the temporal evolution which allows us to appreciate the oscillatory behavior more in detail. An interesting feature of these dynamical states is that the ensemble of the peaks do not oscillate synchronously as a whole, but rather contiguous peaks oscillate out of phase. This behavior is shown in Fig.~\ref{Breather}(c) where the temporal evolution of peaks $p_0$ and $p_1$ [see top profiles in Fig.~\ref{Breather}(a)] is plotted over several periods [see blue and orange lines in Fig.~\ref{Breather}(c)]. The oscillations occur close to the inner part of the structure, and far from the localization boundaries which remain stationary, as one can easily appreciate in Figs.~\ref{Breather}(a)-(b). In Fig.~\ref{Breather}(d) we show the limit cycle described by the projection of the breather dynamics onto the two dimensional phase space defined by Re$[A](p_0)$, and Im$[A](p_0)$.

The breather arises supercritically from H$_{11}$, and therefore with a very small amplitude of oscillation. Decreasing the input pump $S$ this amplitude increases, and the breather persists until reaching its maximal amplitude value at SN$_{11}^l$. Soon after passing that fold the stable LP$_{11}$-breather disappears, and the system jumps to another oscillatory attractor [not shown here].

The blue line in the phase diagram of Fig.~\ref{Phase2} corresponds to H$_{11}$. The LP$_{11}$-breather exists in the gray area between H$_{11}$ and SN$_{11}^{l}$. For the range of parameters studied here H$_{11}$ is always supercritical. The stability analysis reveals that the Hopf bifurcation emerges from a Gavrilov-Guckenhaimer (GG) codimension-two bifurcation \cite{guckenheimer_nonlinear_1983}. At this point  SN$_{11}^{l}$ and H$_{11}$ collide, and therefore this bifurcation is characterized by the three eigenvalues: $\sigma_1=0$, and $\sigma_{2,3}=\pm i\Omega$, with $\Omega>0$  \cite{guckenheimer_nonlinear_1983}. For this reason this point is also known as a Fold-Hopf bifurcation. This scenario is similar to the one appearing in the context of Kerr cavities with anomalous GVD \cite{parra-rivas_dynamics_2014}, and can be related with the presence of temporal chaos \cite{gaspard_local_1993}. However, for the range of parameters studied in this work we have not observed any route leading to chaos.
We have to mention that this oscillatory behavior is not related neither  with the presence of a Hopf instability in the CW state nor with the occurrence of a Turing-Hopf crossing \cite{tlidi_space-time_1999,tzou_homoclinic_2013}, but is a direct consequence of spatial coupling i.e. chromatic dispersion \cite{gomila_excitability_2005}.

\section{Discussion and Conclusions}\label{sec:5}

In this article we have presented a detailed study of the dynamics and bifurcation structure of dissipative LSs arising in dispersive doubly resonant DOPO. To do so we have considered a mean-field model for the description of the signal $A$ and pump $B$ fields, which has been derived in \cite{parra-rivas_frequency_2019}. In this work we have neglected the velocity mismatch between the signal and pump field (i.e. $d=0$).

The LSs studied here form due to the locking of fronts connecting a CW and a Turing pattern in a parameter region where both states are stable. Thus, such state can be seen as a portion of the pattern  embedded in a CW surrounding. We have to point out that these states, commonly known as {\it localized patterns} (LPs), are different to those formed through the locking of DWs connecting two CW states~\cite{parra-rivas_localized_2019}.

LPs undergo a particular bifurcation structure known as {\it homoclinic snaking}: for a fixed detuning $\Delta_1$ the LPs solution branches oscillates back and forth in the driving field intensity $S$ within a well defined snaking or pinning region $[S_-,S_+]$. As a direct consequence of this structure, the system exhibits multi-stability between LPs of different widths. The homoclinic snaking phenomenon is not intrinsic to this system, but generic, arising in many different fields ranging from optics to plant ecology. \cite{woods_heteroclinic_1999,burke_snakes_2007,mercader_convectons_2011,lo_jacono_magnetohydrodynamic_2011,beaume_homoclinic_2011,brena-medina_subcritical_2014,kreilos_fully_2017,zelnik_desertification_2017,gandhi_spatially_2018,parra-rivas_bifurcation_2018}. 
For the set of $(\Delta_1,S)-$parameters studied here, homoclinic snaking is well preserved. Decreasing $|\Delta_1|$ LPs suffer a series of cusp bifurcations, and eventually disappear. Increasing $|\Delta_1|$, although, LPs persist, their bifurcation structure may eventually become more complex.

In most of the physical situations, perfect phase matching is difficult to achieve, and one has to take into account how phase mismatch ($\varrho\neq0$) may perturb the existence and stability of the aforementioned states. We have verified that, for the range of $\varrho$ studied here, LPs are preserved and undergo the same type of bifurcation structure, although their stability may be strongly modified. Indeed for $\varrho<0$ we have found that LPs of different extensions undergo Hopf instabilities, leading to the appearance of LP-breathers, where contiguous peaks oscillate out of phase. This Hopf instability emerges from a Gavrilov-Guckenheimer bifurcation as it was reported in the case of Kerr cavities \cite{parra-rivas_dynamics_2014}. As far as we known this type of dynamical LPs has not been yet reported in the context of dispersive cavities.

We have to mention that, for clarity, two interesting points have been left out this work. First, the LPs presented here and the LS analyzed in \cite{parra-rivas_localized_2019} may coexist and interact, forming much more complex states whose dynamics are yet unexplored. One plausible situation consist in a smooth transition between the homoclinic and collapsed snaking as was described in \cite{zelnik_implications_2018}. The second point focuses on the effect that walk-off may have in the current LPs. On the one hand the walk-off breaks the $x\rightarrow -x$ symmetry, inducing a drift at constant speed, and on the other hand it breaks the homoclinic snaking structure forming a stack of isolas \cite{burke_swift-hohenberg_2009,parra-rivas_third-order_2014}, or more complex configurations \cite{makrides_predicting_2014}. A detailed analysis of these points will appear elsewhere.


\acknowledgments 
We acknowledge support from the FNRS (PPR), and funding from the European Research Council (ERC) under the European Union’s Horizon 2020 research and innovation programme [Grant agreement No. 757800], (FL).

\bibliographystyle{ieeetr}
\bibliography{Dark_OPO}

\end{document}